\documentclass[12pt, preprint]{aastex}

\usepackage{graphicx}
\usepackage{amssymb}
\usepackage{layout}
\usepackage{url}
\usepackage{subfigure}
\usepackage{color}
\slugcomment{Accepted for Publication in {\it The Astrophysical Journal}}

\shorttitle{Polarization Swings Reveal Magnetic Energy Dissipation in Blazars}
\shortauthors{H. Zhang, X. Chen, M. B\"ottcher, F. Guo \& H. Li}

\begin{document}


\title{Polarization Swings Reveal Magnetic Energy Dissipation in Blazars}


\author{Haocheng Zhang\altaffilmark{1,2}, Xuhui Chen\altaffilmark{3,4}, Markus B\"ottcher\altaffilmark{5,1}, 
Fan Guo\altaffilmark{2} and Hui Li\altaffilmark{2}}

\altaffiltext{1}{Astrophysical Institute, Department of Physics and Astronomy, \\
Ohio University, Athens, OH 45701, USA}

\altaffiltext{2}{Theoretical Division, Los Alamos National Laboratory, Los Alamos, NM 87545}

\altaffiltext{3}{Institute of Physics and Astronomy, University of Potsdam, 14476 Potsdam-Golm, Germany}

\altaffiltext{4}{DESY, Platanenallee 6, 15738 Zeuthen, Germany}

\altaffiltext{5}{Centre for Space Research, North-West University, Potchefstroom,
2520, South Africa}

\begin{abstract}
The polarization signatures of the blazar emissions are known to be highly variable. 
In addition to small fluctuations of the polarization angle around a mean value, sometimes 
large ($\gtrsim 180^{\circ}$) polarization angle swings are observed. We suggest that such p
henomena can be interpreted as arising from light-travel-time effects within an underlying 
axisymmetric emission region. We present the first simultaneous fitting of the multi-wavelength 
spectrum, variability and time-dependent polarization features of a correlated optical and 
gamma-ray flaring event of the prominent blazar 3C279, which was accompanied by a drastic 
change of its polarization signatures. This unprecedented combination of spectral, variability, 
and polarization information in a coherent physical model allows us to place stringent constraints 
on the particle acceleration and magnetic-field topology in the relativistic jet of a blazar, 
strongly favoring a scenario in which magnetic energy dissipation is the primary driver of the 
flare event.
\end{abstract}

\section{Introduction}

Blazars are the most violent active galactic nuclei. They are known to emit nonthermal-dominated 
radiation from radio frequencies to $\gamma$-rays, with strong variability across the entire
spectrum \citep[e.g.,][]{Aharonian07}. Such emission likely originates from a relativistic jet 
that is directed close to our line of sight. The blazar spectral-energy-distribution (SED) has 
two components. It is generally acknowledged that the nonthermal radio through optical-UV emission 
is synchrotron radiation from ultrarelativistic electrons. In leptonic models, the high-energy component, 
from X-rays to $\gamma$-rays, is usually interpreted as resulting from inverse Compton scattering by 
the same nonthermal electron population of soft seed photons. Seed photons can either come from the 
synchrotron component itself \citep[synchrotron-self Compton = SSC; e.g.,][]{Maraschi92,Marscher85}, 
or from external photon fields originaiting in the broad line region and/or a dusty torus \citep[external 
Compton = EC; e.g.,][]{Dermer92,Sikora94}. However, a hadronic origin of the high-energy component 
cannot be ruled out \citep[e.g.,][]{Mannheim92,Mucke01,Boettcher13}. Recently, \cite{Zhang13} have 
demonstrated that the high-energy polarization signatures can serve as a powerful diagnostic between 
the leptonic and hadronic processes.

Despite intensive observational and theoretical efforts, the particle acceleration mechanism, the energy 
source for flaring activity, and the inner-jet physical conditions, such as the magnetic field topology, 
are not well understood. On the observational side, the synchrotron component is known to be polarized, 
with polarization degrees ranging from a few to tens of percent, in agreement with a synchrotron origin 
in a partially ordered magnetic field. Several authors \citep[e.g.,][]{Lyutikov05,Pushkarev05} have 
shown that the observed polarization degree and angle may reveal a helical magnetic field structure. 
Both the polarization degree and angle are known to be highly variable. Large ($\ge 180^{\circ}$) 
polarization angle swings have been frequently observed, and some recent observations 
\citep{Marscher08,Marscher10,Abdo10} have shown that they are sometimes accompanied by 
simultaneous optical and $\gamma$-ray flaring activities. When the polarization angle is 
not rotating, small fluctuations of the polarization signatures around a relatively stable 
mean value are often observed \citep[e.g.][]{Aleksic14,Morozova14}. Time-symmetric polarization 
profiles are observed in both cases, in particular a complete $\sim180^{\circ}$ polarization angle 
swing is often accompanied by a drop of the polarization degree to nearly zero followed by a recovery 
back to its value before the PA rotation (see the above references). On the theory side, the general 
formalism for synchrotron polarization is well understood \citep[e.g.,][]{Westfold59}. Nevertheless, 
the spectrum, variability and polarization signatures have never been combined into one coherent 
physical model to understand the inner-jet physics. Models for synchrotron polarization necessarily 
take into account the magnetic field topology and polarization-dependent synchrotron emissivity, 
but often apply simple, time-independent, power-law electron populations, and usually ignore the 
generation of high energy emission \citep[e.g.,][]{Lyutikov05}. Thus they cannot produce full 
broadband spectral-energy-distributions (SED) and variability. On the other hand, models presenting 
detailed calculations of broadband emission and variability typically assume a chaotic magnetic 
field, and ignore any polarization-dependent emissivity and polarization information. 
Additionally, previous emission models hardly provide any constraints on the particle 
acceleration mechanisms and the energy source. In a recent paper, \cite{Zhang14} (hereafter 
Paper I) have presented a general parameter study as a first step to combine the full SED, 
variability and polarization signatures in a single physical model. This paper has demonstrated 
that polarization-angle swings, correlated with high-energy flaring activity, only require a 
dominant helical magnetic field structure in a straight jet, but no non-axisymmetric jet 
features, such as bends or helical flow streamlines.

In this paper, we will investigate additional geometric effects, namely the shape of the emission 
region and the magnetic field topology, on the synchrotron polarization signatures and multiwavelength 
variability. We find that the apparently time-symmetric polarization profiles can be obtained by an 
axisymmetric emission region along with full consideration of the light-travel-time effects. We will 
then apply the results obtained here and in Paper I to present the first simultaneous fitting of 
snapshot broadband SEDs, light curves in various frequency bands, and time-dependent polarization 
signatures, from a correlated optical and $\gamma$-ray flaring event of the prominent blazar 3C279, 
which was accompanied by a drastic change of its polarization features. This unprecedented fitting 
combination in one coherent model allows us to place stringent constraints on the particle acceleration 
and its energy source, as well as the magnetic field structure and its variation, in the inner jet of 
the blazar. Our results strongly favor a scenario in which magnetic energy dissipation is the primary 
driver of the flare event. We will describe our model setup in Section 2, present a general study of 
the geometric effects in Section 3, fit the 3C279 flare event in Section 4, and discuss the results 
in Section 5.

\section{Model Setup}

Our setup is designed to highlight the combined geometric and light-travel-time effects with the 
simplest physical assumptions. The underlying model considered here assumes a cylindrical emission 
region which travels on a straight trajectory defined by the jet boundary and encounters a flat 
stationary disturbance. The entire region is pervaded by a predominantly helical magnetic field, 
with possibly an additional turbulent component. The origin of the disturbance could be a shock, 
which will convert bulk kinetic energy into random motion of relativistic particles to generate 
a flare; but it may also may also be associated with other mechanisms such as magnetic reconnection, 
which will dissipate magnetic energy to accelerate particles (\cite{Guo14}, see Section 4). In the 
comoving frame of the emission region, the disturbance will travel through the emission region, 
and temporarily change the plasma conditions at its location,primarily the magnetic field structure 
and the nonthermal particle distribution through a modification of the acceleration time scale, and 
hence generate a flare. As the disturbance leaves a given point in the emission region, the plasma 
conditions will recover back to the initial state. Most importantly, due to the light-travel-time 
effects (LTTEs),  the observer will receive signatures of flaring activities induced by the disturbance, 
from points at different longitudinal distances along the jet at the same time. Owing to the axisymmetry 
of the emission region, this leads to time-symmetric polarization profiles (see Sections 3 and 4).

\begin{figure}[ht]
\centering
\includegraphics[width=7.5cm]{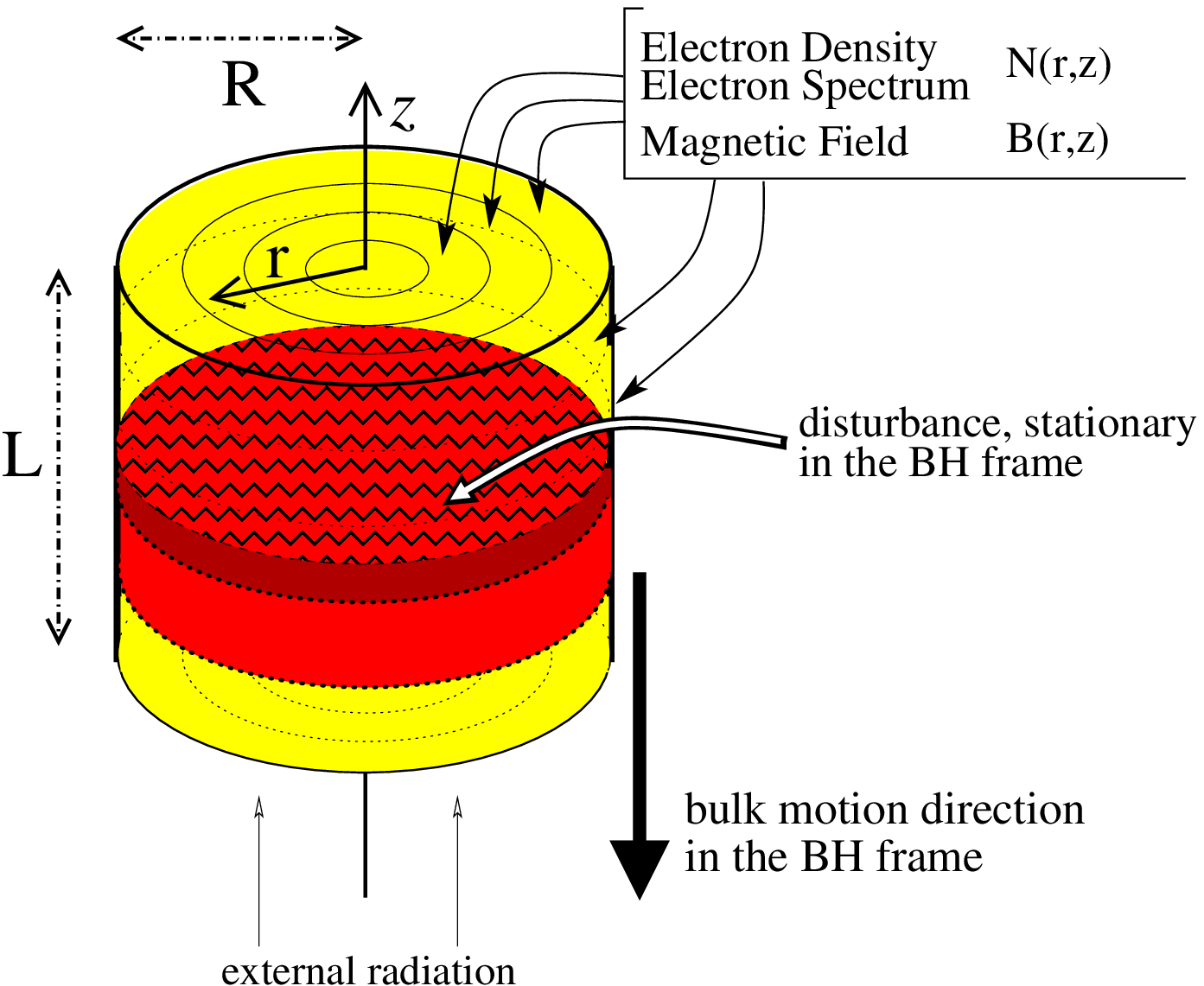}
\hspace{0.5cm}
\includegraphics[width=6.5cm]{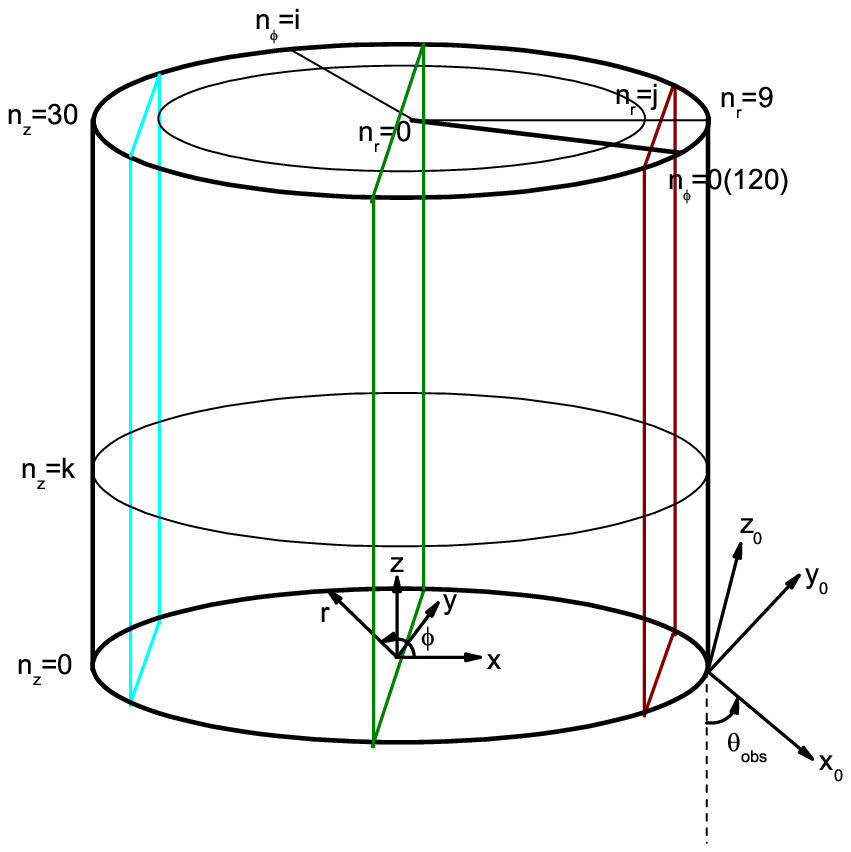}
\caption{{\it Left:} Sketch of the geometry used in the MCFP code, adapted from \cite{Chen12}.
{\it Right:} Sketch of the geometry (in the co-moving frame) of 3DPol. The model uses cylindrical
coordinates, ($r$, $\phi$, $z$), with $n_r$, $n_{\phi}$, $n_z$ being the number of zones in the
respective directions. We define a corresponding Cartesian coordinate system ($x, y, z$) where
$z$ is along the axis of the cylinder and the $x$-axis is along the projection of the LOS onto the
plane perpendicular to $z$. The Cartesian coordinates ($x_0, y_0, z_0$) are defined so that
$x_0$ is along the LOS and $z_0$ is the projection of the cylindrical axis onto the plane of the
sky. Both Cartesian coordinate systems are in the co-moving frame of the emission
region. $\theta_{\rm obs}$ is the observing angle between $x_0$ and $z$. Consequently, if
$\theta_{\rm obs} = 90^{\circ}$, ($x, y, z$) = ($x_0, y_0, z_0$). The cyan, dark-green and maroon
regions represent far-side (left), middle and near-side (right) zones, respectively.
\label{setup}}
\end{figure}

The above model is simulated by the combination of the two-dimensional Monte-Carlo/Fokker-Planck 
radiation transfer (MCFP) code developed by \cite{Chen12,Chen14} and the 3D multi-zone synchrotron 
polarization ray-tracing code developed by \cite{Zhang14}. Fig. \ref{setup} shows the model and code 
setup. The MCFP code considers a cylindrical emission region, assumes axisymmetry for all physical 
properties, and divides the region evenly into multiple zones in the radial and longitudinal directions. 
The electron distribution in each zone evolves in time according to a locally isotropic Fokker-Planck 
equation. The disturbance is also a cylindrical region, which will instantaneously modify the physical 
conditions of the zones at its current location, hence modifying the Fokker-Planck evolution. When it 
moves out of a zone, the physical conditions will instantaneously be restored to the initial state. 
Here, the physical meaning of an ``instantaneous'' change is that its time scale is less than one 
time step of the simulation. The code will then calculate the time-dependent emissivities based on 
the particle distribution at each time step, and apply a Monte-Carlo method to trace the photons 
and handle the Compton scattering. Therefore, all LTTEs are naturally included. All calculations 
are performed in the comoving frame of the emission region, and the resulting fluxes are properly 
transformed to the observer's frame at the end of the simulation.

The 3DPol code is focused on the synchrotron polarization signatures. It uses the same geometry 
and physical conditions, and the particle distributions as calculated by the MCFP code. However, 
since the polarization requires a 3D description, the emission region is further divided evenly 
into multiple zones in the $\phi$ direction, but the axisymmetry of all parameters is still kept. 
Since the environment of the emission region is generally considered as optically thin, and our 
treatment of the synchrotron polarization focuses on the high-frequency radio through optical/UV 
range, synchrotron-self absorption and Faraday rotation effects are neglected in this study. As 
in the MCFP code, all calculations are performed in the comoving frame. In each zone, we project 
the magnetic field onto the plane of sky in the comoving frame, and use the time-dependent electron 
population generated by the MCFP code to evaluate the polarization degree ($PD$) at various frequencies. 
Since the net electric vector is perpendicular to the projected magnetic field direction, the electric 
vector position angle, also known as the polarization angle ($PA$), is easily obtained. In this way, 
we obtain the Stokes parameters (without normalization) at various frequencies at every time step in 
each zone via
\begin{equation}
(I,Q,U)_{\nu} = L_{\nu}*(1,\Pi_{\nu}\cos2\theta_E,\Pi_{\nu}\sin2\theta_E)
\end{equation}
where $L_{\nu}$ and $\Pi_{\nu}$ are the spectral luminosity and the $PD$ at frequency $\nu$, respectively,
and $\theta_E$ is the electric vector position angle for that zone. The code then employs ray-tracing to 
calculate the relative time delay to the observer for each zone, in order to take full account of the 
LTTEs. Since the emissions from different zones are incoherent, the total Stokes parameters can be 
obtained by linear addition of the Stokes parameters from each zone for emissions reaching the observer 
at the same time. Finally, as in the MCFP code, all emissions are properly Lorentz transformed (boosted) 
to obtain the luminosity and polarization information in the observer's frame.

Due to the relativistic aberration, even though we are observing blazars nearly along the jet
in the observer's frame (typically, $\theta^{\ast}_{\rm obs} \sim 1/\Gamma$, where $\Gamma$ is
the bulk Lorentz factor of the outflow along the jet), the angle $\theta_{\rm obs}$ between LOS
and the jet axis in the comoving frame it is likely much larger. Specifically, if $\theta^{\ast}_{\rm obs}
= 1/\Gamma$, then $\theta_{\rm obs} = 90^{\circ}$. Paper I has shown that the polarization signatures 
are only weakly dependent on the exact value of $\theta_{\rm obs}$, thus in all of the following 
simulations, we choose $\theta_{\rm obs} = 90^{\circ}$ in the comoving frame, and hence, the Doppler 
factor $\delta \equiv \left( \Gamma \, [1 - \beta_{\Gamma} \, \cos\theta_{\rm obs}^{\ast}] \right)^{-1} 
= \Gamma$.

\section{Geometric Effects\label{section3}}

In this section, we use the quasar PKS 1510-089 as an example to investigate a number of generic 
flaring scenarios as case studies. PKS~1510-089 is
an Flat-Spectrum Radio Quasar (FSRQ). Its flaring activity in March 2009
was well covered by multi-band monitoring in the infrared, optical, X-rays, and $\gamma$-rays
\citep{abdo10b,dammando11,Marscher10}. \cite{Chen12} presented model fits to snap-shot SEDs
and light curves, in which they found that the most favorable scenario is an external Compton (EC) model
with the external radiation field originating from a dusty torus.

The purpose of this section is to study geometric effects on the polarization signatures, in particular 
the $PA$ variability.
The geometries are chosen in such a way that the first two cases correspond to $180^{\circ}$ $PA$ swings, 
while the other two cases are aimed to study to study small-scale PA fluctuations.
In order to facilitate a direct comparison with paper I, we choose the same quiescent state parameters,
similar to those used in \cite{Chen12}. In addition,
we keep the assumption that the disturbance is due to a shock.
However, the shock parameters and the geometry may vary. Also, for the purpose of these generic case studies, 
we do not introduce any turbulent
magnetic field contribution. Since both the MCFP and 3DPol codes are time-dependent, we allow for an initial 
period for the electrons and the photons to reach equilibrium,
before we introduce the parameter disturbance produced by the shock.
In all our results the plots illustrate the results after this equilibrium has been reached. As the flaring 
activity
exhibits different characteristics in duration and in strength for different cases, we define
similar phases in the flare development for the purpose of a direct comparison. These phases
correspond approximately to the pre-flare, early flare, flare peak, late flare, and post-flare states.

We define the $PA$ in our simulations as follows. $PA=0$ corresponds to the electric field
vector being parallel to the projection of the cylindrical axis on the plane of sky. An increasing 
$PA$ corresponds
to counter-clockwise rotation with respect to the LOS, to $180^{\circ}$ when it is anti-parallel
to the projected cylindrical axis (which is equivalent to $0$ due to the $180^{\circ}$ ambiguity).
Based on the above definition, the Stokes parameters normalized by luminosity for one zone with
its projected magnetic field directed in the range of $0$ -- $45^{\circ}$
is in the form of $(1,-\vert q \vert, \vert u \vert)$, in the range $45^{\circ}$ -- $90^{\circ}$ 
it is $(1, \vert q \vert, \vert u \vert)$, for $-45^{\circ}$ -- $0$ it is $(1,-\vert q \vert, -\vert u \vert)$,
and for $-90^{\circ}$ -- $45^{\circ}$ it is $(1, \vert q \vert, -\vert u \vert)$. This convention will 
be frequently used in the following text.
All results are shown in the observer's frame.

Table 1 lists some key parameters. The emission region is a cylinder of length $Z$ and radius $R$, 
traveling at a bulk Lorentz factor of $\Gamma$ in the observer's frame, when it encounters a flat, 
uniform, stationary disturbance of length $L$ and radius $A$. The Fokker-Planck equation takes into 
account the acceleration time-scale $t_{\rm acc}$ and escape time-scale $t_{\rm esc}$, both in units 
of $Z/c$; however, the total number of electrons is conserved, hence the escaping electrons will be 
balanced by thermal electrons picked up from the background. In the quiescent state the emission 
region is pervaded by a helical magnetic field $B_H$ oriented at a pitch angle $\theta_B$, and 
possibly a turbulent contribution, where the total magnetic-field strength is $B$. However, in 
this geometric effect study, we do not introduce any turbulence, therefore $B_H=B$. Initially a 
power-law distribution of nonthermal electrons with a power-index $p$ and minimum and maximum 
energies $\gamma_{\rm min}$ and $\gamma_{\rm max}$, respectively, with density $n_e$, will have 
evolved to an equilibrium according to the Fokker-Planck equation in the emission region before 
the interaction with the disturbance. The disturbance will change the layers in the emission 
region at its location into an active state. In cases 1, 2 and 4, the disturbance, which is in 
the form of a shock, will compress the local magnetic field, so that the magnetic field strength 
will be amplified by a factor of $B^s/B$, and the pitch angle will change to $\theta^s_B$. In 
case 3, the shock will instead shorten the acceleration time scale by a factor of $t^s_{\rm acc}/t_{\rm acc}$, 
so as to increase the acceleration efficiency.

\begin{table}[ht]
\scriptsize
\parbox{1.0\linewidth}{
\centering
\begin{tabular}{|l|c|}\hline
Parameters                                          & General Properties     \\ \hline
Bulk Lorentz factor $\Gamma$                        & $15.0$                 \\ \hline
Total (helical) magnetic field $B$ $(G)$            & $0.2$                  \\ \hline
Initial electron density $n_e$ $(10^{2}cm^{-3})$    & $7.37$                 \\ \hline
Initial electron minimum energy $\gamma_{\rm min}$      & $50$                   \\ \hline
Initial electron maximum energy $\gamma_{\rm max}$      & $20000$                \\ \hline
Initial electron spectral index $p$                 & $3.2$                  \\ \hline
Electron acceleration time-scale $t_{\rm acc}$ $(Z/c)$  & $0.09$                 \\ \hline
Electron escape time-scale $t_{\rm esc}$ $(Z/c)$        & $0.015$                \\ \hline
Orientation of LOS $\theta_{\rm obs}$ $(^{\circ})$      & $90$                   \\ \hline
\multicolumn{2}{c}{}\\
\end{tabular}}
\parbox{1.0\linewidth}{
\centering
\begin{tabular}{|l|c|c|c|c|}\hline
Parameters                                          & Case 1            & Case 2            & Case 3         & Case 4         \\ \hline
Helical pitch angle $\theta_{B}$ $(^{\circ})$       & $45$              & $45$              & $65$           & $45$           \\ \hline
Length of the emission region $Z$ $(10^{16}cm)$     & $8.0$             & $18.0$            & $18.0$         & $8.0$          \\ \hline
Radius of the emission region $R$ $(10^{16}cm)$     & $6.0$             & $4.0$             & $4.0$          & $6.0$          \\ \hline
Length of the disturbance $L$ $(10^{16}cm)$         & $0.8$             & $1.8$             & $1.8$          & $0.8$          \\ \hline
Radius of the disturbance $A$ $(10^{16}cm)$         & $6.0$             & $4.0$             & $4.0$          & $1.778$        \\ \hline
$B^s/B$                                             & $\sqrt{3}+1$      & $\sqrt{3}+1$      & $--$           & $\sqrt{3}+1$   \\ \hline
$\theta_{B}^s$ $(^{\circ})$                         & $75$              & $75$              & $--$           & $75$           \\ \hline
$t_{\rm acc}^s/t_{\rm acc}$                                 & $--$              & $--$              & $1/2.5$        & $--$           \\ \hline
\end{tabular}}
\caption{Summary of model parameters. {\it Top:} General properties valid for all cases. Notice that 
$\gamma_{\rm min}$, $\gamma_{\rm max}$
and $p$ describe the injection of nonthermal particles and the electron distribution will reach an 
equilibrium before interacting with the disturbance. {\it Bottom:} Parameters for each case.
The $s$-superscript indicates the parameters during the presence of the shock. The parameter $B^s/B$ 
is the magnetic-field amplification factor,
$t_{\rm acc}^s/t_{\rm acc}$ is the acceleration time scale shortening factor, and
$\theta_{B}^s$ is the magnetic-field pitch angle in the shocked region.}
\end{table}

\subsection{B-Field Compression, $Z<2R$\label{section31}}

In this scenario, the shock will instantaneously increase the toroidal magnetic field component at its
location, probably by compression, so that it will increase the total magnetic field strength and change its
orientation in the affected zones. The new magnetic field will be kept until the shock moves out of that zone,
when it instantaneously reverts back to its original (quiescent) strength and orientation,
which can be attributed to dissipation. Paper I has considered a very extreme case, where the shock 
increases the toroidal component by a factor of ten,
so that the total magnetic field is increased by approximately a factor of seven, and $\theta_B$ increases
from initially $45^{\circ}$ to approximately $84^{\circ}$. Such strong alteration in the magnetic field
results in a drastic change in the polarization signatures; however it will require a huge amount of 
energy conversion
within small time windows when the disturbance is turned on and off, which is unlikely to happen in practice.
Here we investigate a case with a moderate change in the magnetic field, where $\theta_B$ of the total magnetic
field increases from $45^{\circ}$ to $75^{\circ}$. To mimic a spherical volume for the emission region, 
we choose $R:Z=1:\frac{4}{3}$,
so that $Z < 2 R$, and the disturbance will occupy the entire layer, i.e., $A=R$ (hereafter Geometry I).

\begin{figure}[ht]
\centering
\includegraphics[width=15cm]{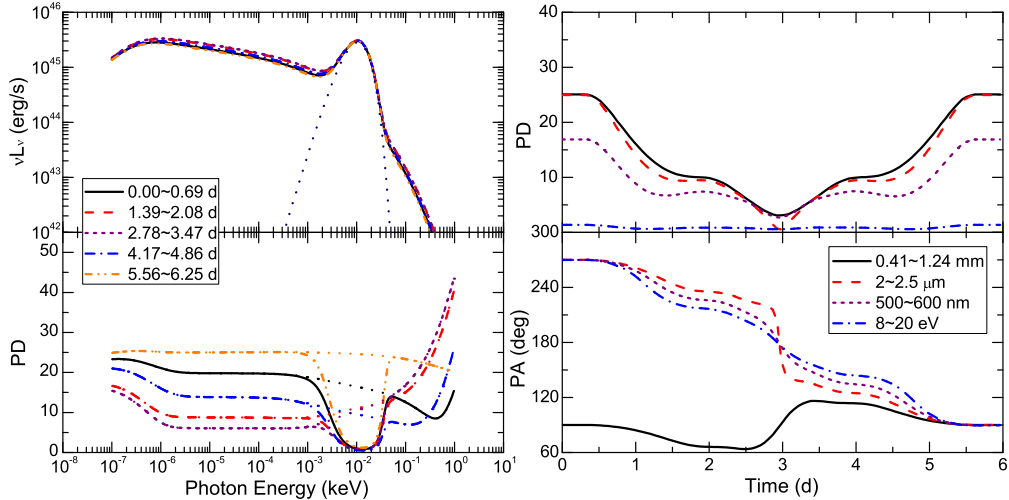}
\caption{Case 1: A moderate change of magnetic field in Geometry I.
{\it Upper left:} Snap-shot synchrotron SEDs, including the external photon field.
SEDs are chosen at approximately the pre-flare (black solid), early flare (red dashed),
peak (purple short dashed), late flare (blue dash-dotted) and back to equilibrium at the 
post-flare (orange dash-dot-dotted) states,
with the dotted line for the external photon field contribution.
All SEDs are chosen with the same time bin size.
{\it Lower left:} $PD$ vs. photon energy for the same time bins as the SEDs in the top panel, 
with the external photon contamination considered,
where dotted lines represent the $PD$ without the contamination.
{\it Upper right:} $PD$ vs. time with external contamination, at radio (black solid),
infrared (red dashed), optical V band (purple short dashed), UV (blue dash-dotted).
{\it Lower right:} $PA$ vs. time for the energy bands as in the top panel.
\label{Case11}}
\end{figure}

Since the total magnetic field is not increased much, we observe in Fig. \ref{Case11} (upper left) that
the synchrotron flux exhibits only a small flare. It is also obvious that the $PD$ is photon-energy 
dependent (Fig. \ref{Case11} lower left) since PKS~1510-089 has a visible
external thermal component emerging at optical-UV frequencies, which diminishes the observed polarization
percentage, especially in the UV. The dotted lines in Fig. \ref{Case11} (lower left) show the intrinsic
$PD$ without the thermal contamination. The spectral curvature of the realistic electron
spectrum from the detailed simulation based on the Fokker-Planck equation leads to an additional 
frequency dependence of the $PD$. The features above
$\sim 100$~eV are due to the electron spectral cutoff. For the time dependence of the polarization 
signatures, the $PD$ gradually decreases to nearly zero at the flare peak, then recovers back to its 
initial state at the
end of the flare. Moreover, the $PA$ shows a continuous $180^{\circ}$ swing, except for the radio/mm 
polarization,
which is due to the extremely small flare amplitude at low frequencies. As is elaborated below and 
in Paper I,
such polarization variability features can be explained as the combined effect of electron evolution 
and LTTEs.

\begin{figure}[ht]
\centering
\includegraphics[width=15cm]{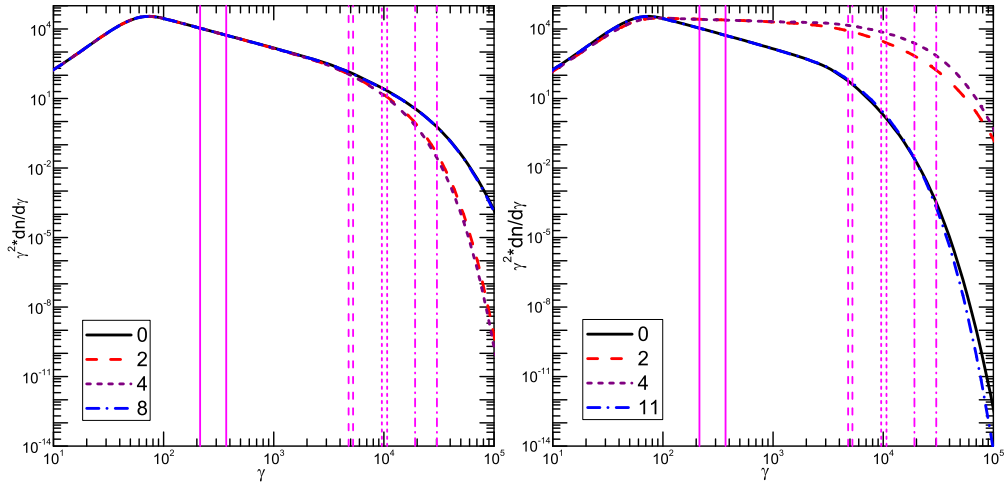}
\caption{Electron spectra at different time steps for a moderate change in the magnetic field 
({\it left}) and a shortening of the acceleration time scale ({\it right}).
$0$: Pre-flare equilibrium;
$2$: central shock position; $4$: shock just leaves the emission region; $8(11)$: post-shock equilibrium.
The regions between the magenta vertical lines represent
the electron energies that correspond to the photon energies we choose in the polarization vs. time plots:
Solid corresponds to radio, dashed to infrared, short-dashed to optical, and dashed-dotted to UV.
\label{ElectronSpectrum}}
\end{figure}

\begin{figure}[ht]
\centering
\includegraphics[width=0.9\textwidth]{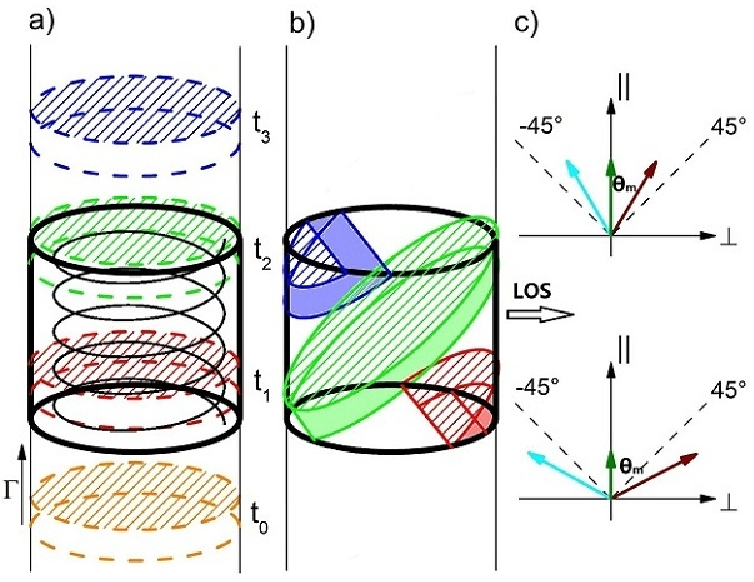}
\caption{a) Sketch of the interaction between the emission region and the disturbance in the 
comoving frame of the emission region, at different epochs. The emission region is pervaded 
by a helical magnetic field and a turbulent component (only the helical component is sketched). 
The disturbance is stationary in the observer's frame, but in the comoving frame of the emission 
region, the disturbance is then moving up with Lorentz factor $\Gamma$. The part of the disturbance 
that first encounters the emission region (shaded ellipses, the ``front'') is the location of the 
injection of relativistic particles. The orange, red, green and blue regions refer to the locations 
of the disturbance before the flare ($t_0$), the rising phase ($t_1$), peak ($t_2$) and declining 
phase ($t_3$), respectively. b) The red, green and blue shapes indicate the shape and location of 
the flaring region, corresponding to the disturbance at $t_1\sim t_3$, respectively, observed 
simultaneously, taking into account the LTTEs. c) Sketch of the projection of the helical magnetic 
field onto the plane of sky in the comoving frame. The upper panel illustrates the quiescent state, 
the lower panel the active state. The cyan, dark-green and maroon arrows represent the left side, 
center and the right side of the emission region shown in a), corresponding to the color regions 
in Fig. \ref{setup}, respectively. $\parallel (\perp)$ denotes components parallel (perpendicular) 
to the bulk motion direction. The two dashed lines indicate $\pm 45^{\circ}$.
\label{Sketch}}
\end{figure}

Initially, the directions of the projected
magnetic field will cover the range of $-45^{\circ}$ to $45^{\circ}$.
Due to the axisymmetry of the initial conditions, the initial $\vert u \vert$ component will be 
balanced out, hence the initial Stokes vector is $(1, - \vert q \vert, 0)$,
leaving $PA$ at $270^{\circ}$ (equivalent to $90^{\circ}$).
When the shock moves into the emission region, it will increase the toroidal component of the 
magnetic field at its location,
resulting in stronger emission and enhanced synchrotron cooling (Fig. \ref{ElectronSpectrum} left). 
Additionally, during the presence of the disturbance, the active region has $\theta_B=75^{\circ}$.
The combined effect is that the active region will have a larger contribution to the polarization, 
with the Stokes parameters generally in the form of
$(1, \vert q \vert, \vert u \vert)$ or $(1, \vert q \vert, -\vert u \vert)$.
Due to LTTEs, although the shock is flat in the co-moving frame,
the active region at equal photon-arrival times will be distorted, as shown in Fig. \ref{Sketch}a, b.
In the early flare phase, only the region facing the observer (red in Fig. \ref{Sketch}) contributes 
to the flare,
where the initial Stokes parameters in the form of $(1,-\vert q \vert, \vert u \vert)$ will be 
replaced by the active state $(1, \vert q \vert, \vert u \vert)$.
Therefore the initial $PA=270^{\circ}$ will gradually move to $225^{\circ}$, and the $PD$ will 
decrease at the same time. When the flare gradually rises up to its peak,
emission from the far side of the cylinder can be seen, which possess negative $u$ component.
This will diminish the positive $u$ from the right side of the cylinder, meanwhile the initial negative $q$
has mostly be canceled out. Consequently the $PD$ will continue to drop, and the $PA$ will rotate 
from $225^{\circ}$
to $180^{\circ}$. When the flare reaches its maximum (green in Fig. \ref{Sketch}b), for a short period
the shock-enhanced fluxes from both sides of the cylinder will be comparable. As a result,
the $u$ component will be canceled out, leaving a positive $q$ in the active region and a 
negative $q$ in the quiescent region.
If the shock is adequately strong, the polarization contributed by the active region will 
be higher, leading to a positive $q$ for this short period. Therefore, the $PD$ drops to nearly
zero for all bands, while the $PA$ reaches $180^{\circ}$ for infrared, optical and UV. However, 
the $PA$ rotates back to the initial
$90^{\circ}$ for the radio, as its flare amplitude is not strong enough to dominate over the 
quiescent emission. After the peak, the flaring region moves to the far side,
so that the polarization signatures gradually revert back to the initial state in time-symmetric patterns.

\subsection{B-Field Compression, $Z>2R$}

In order to isolate the dependence of the polarization signatures on the geometry of the emission 
region, we studied a scenario in which we kept all the parameters exactly the same as in the
previous case, except for the size of the emission region. We choose an emission region with the 
same volume, but now $R:Z=\frac{2}{3}:3$
(hereafter Geometry II). The general electron spectrum is similar to the previous case, except 
minor differences due to the electron cooling rate, escape rate, etc. introduced by the different geometry.
However, we notice that the flare amplitude is slightly increased, thus the radio polarization 
signatures now behave similarly to the other bands.
The major difference in this case is that in this scenario, the $PD$ goes through two minima near 
zero with an elevated plateau inbetween. The $PA$ exhibits a similar step-like pattern with a 
plateau at $180^{\circ}$, although, in total, it still completes a full $180^{\circ}$ swing. 
The reason is that unlike in the previous case, during the peak, the entire equal-photon-arrival-time 
ellipse is maintained within the emission region for a finite amount of time, during which the 
polarization signatures remain unchanged.

\subsection{Acceleration Efficiency, $Z>2R$}

We investigate one additional scenario where the shock leads to more efficient particle acceleration, 
by instantaneously
shortening the acceleration time scale at its location. In this case, we choose the initial toroidal 
component slightly dominant, with the application of Geometry II
(we have shown one case with Geometry I in Paper I). Due to the more efficient acceleration,
the electrons are accelerated to higher energy, and the spectrum becomes harder
(Fig. \ref{ElectronSpectrum} right). Therefore, the synchrotron component exhibits a considerable flare
at higher energies. However, at lower energies the electron
spectrum stays almost unchanged. Since the magnetic field remains unchanged, the radio band shows no time
dependency of either flux or polarization signatures.

At the beginning of the flare, the luminosity of the active region is drastically enhanced;
as a result, the observed $PD$ will mostly be attributed
to the red region in Fig. \ref{Sketch} (right). In this small region,
the magnetic field is well ordered, consequently we observe that $PD$ can shoot up to over
$60\%$; meanwhile $PA$ quickly rotates to about $210^{\circ}$, which represent the $\theta_B$ in this region.
Similar to the previous case, during the flare peak, the entire emission
region becomes axisymmetric, therefore $PA$ rotates back, and both $PD$
and $PA$ exhibit a step phase. After the peak, $PD$ and $PA$ revert back to the initial state in 
time-symmetric patterns.

\begin{figure}[ht]
\centering
\includegraphics[width=15cm]{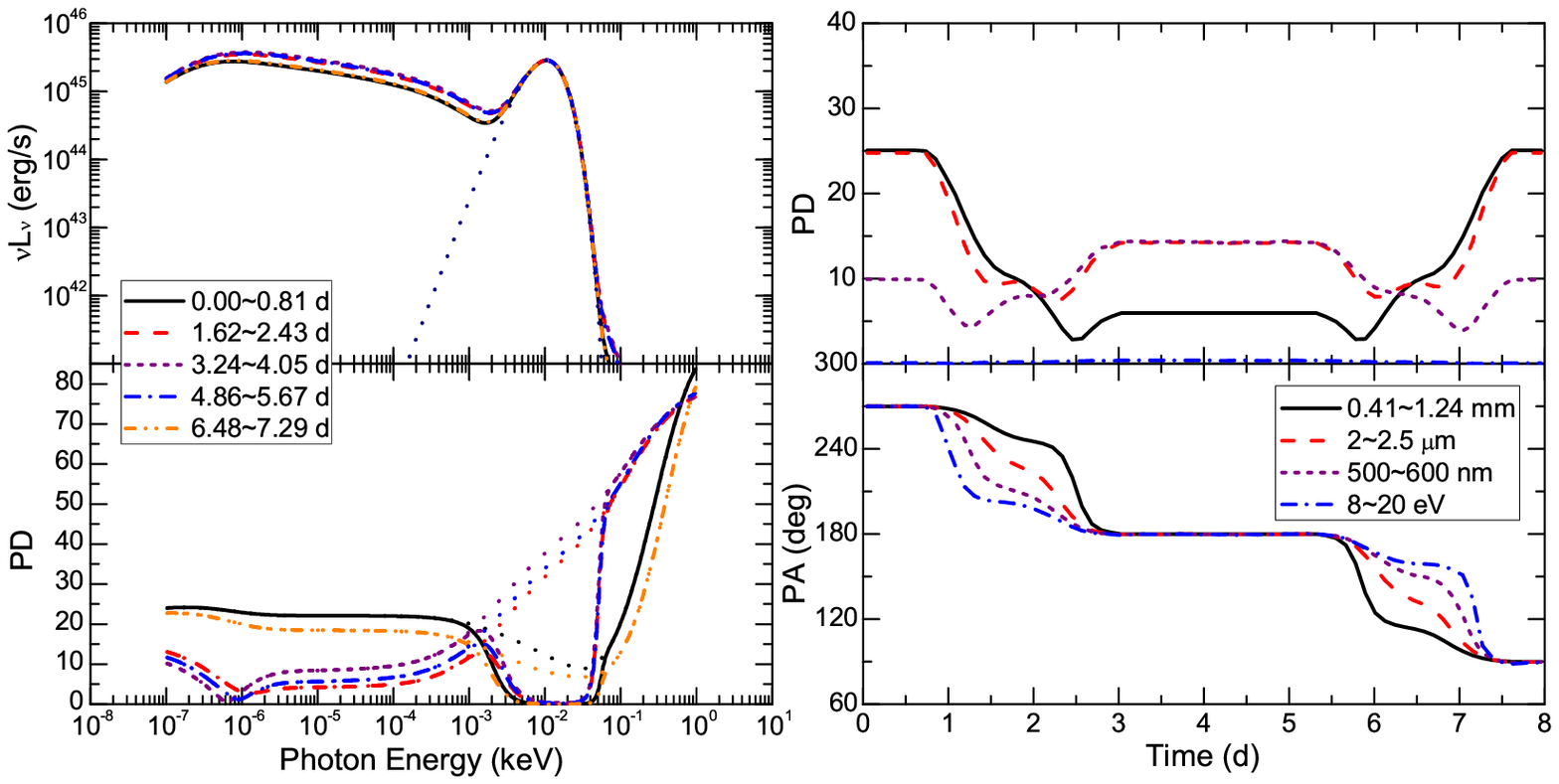}
\caption{Case 2: A moderate change in magnetic field in Geometry II.
Panels and line styles are as in Fig. \ref{Case11}.
\label{Case12}}
\end{figure}

\begin{figure}[ht]
\centering
\includegraphics[width=15cm]{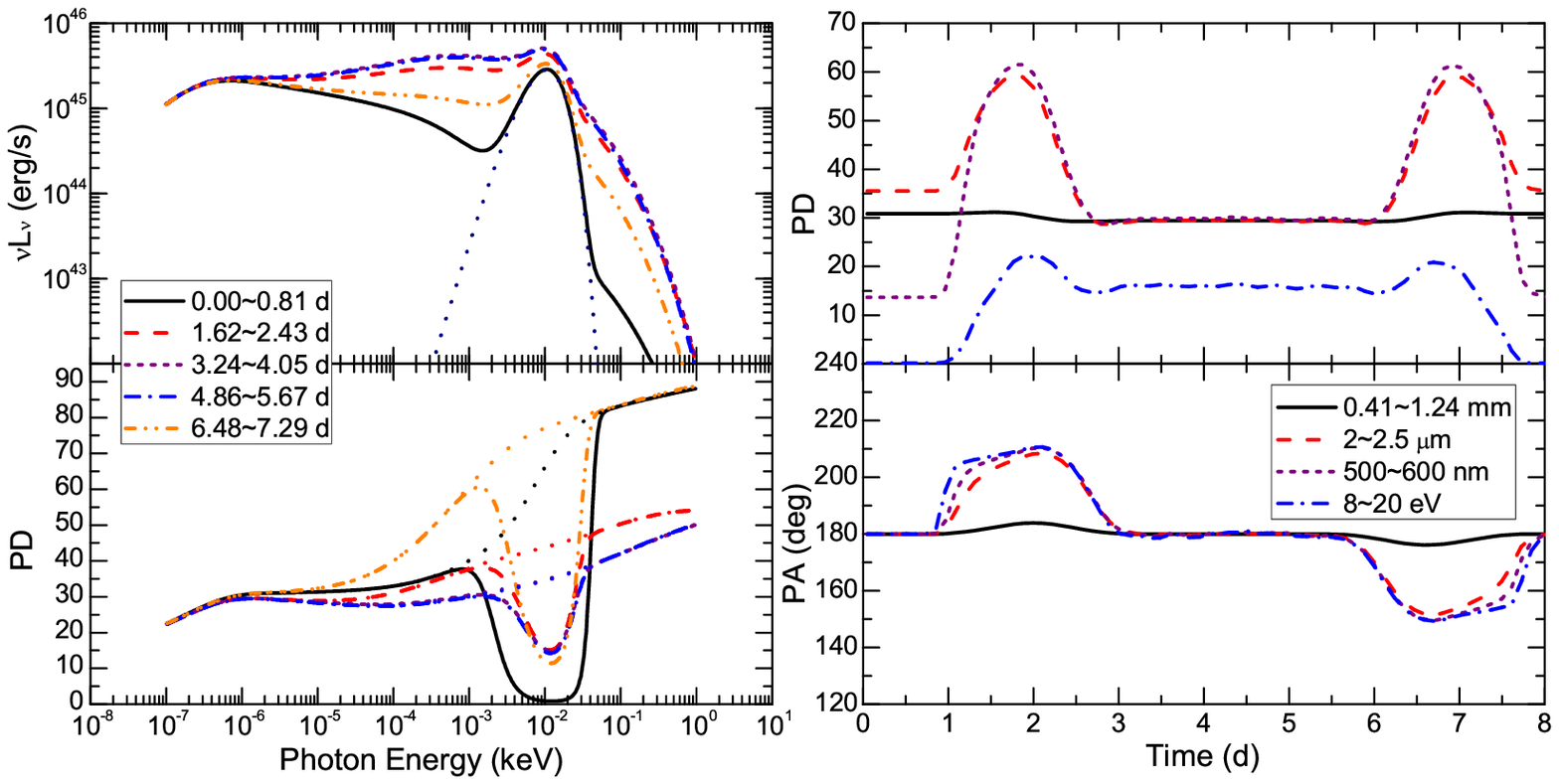}
\caption{Case 3: Shortened acceleration timescale in Geometry II.
Panels and line styles are as in Fig. \ref{Case11}.
\label{Case21}}
\end{figure}

\subsection{B-Field Compression, Localized shock}

\begin{figure}[ht]
\centering
\includegraphics[width=15cm]{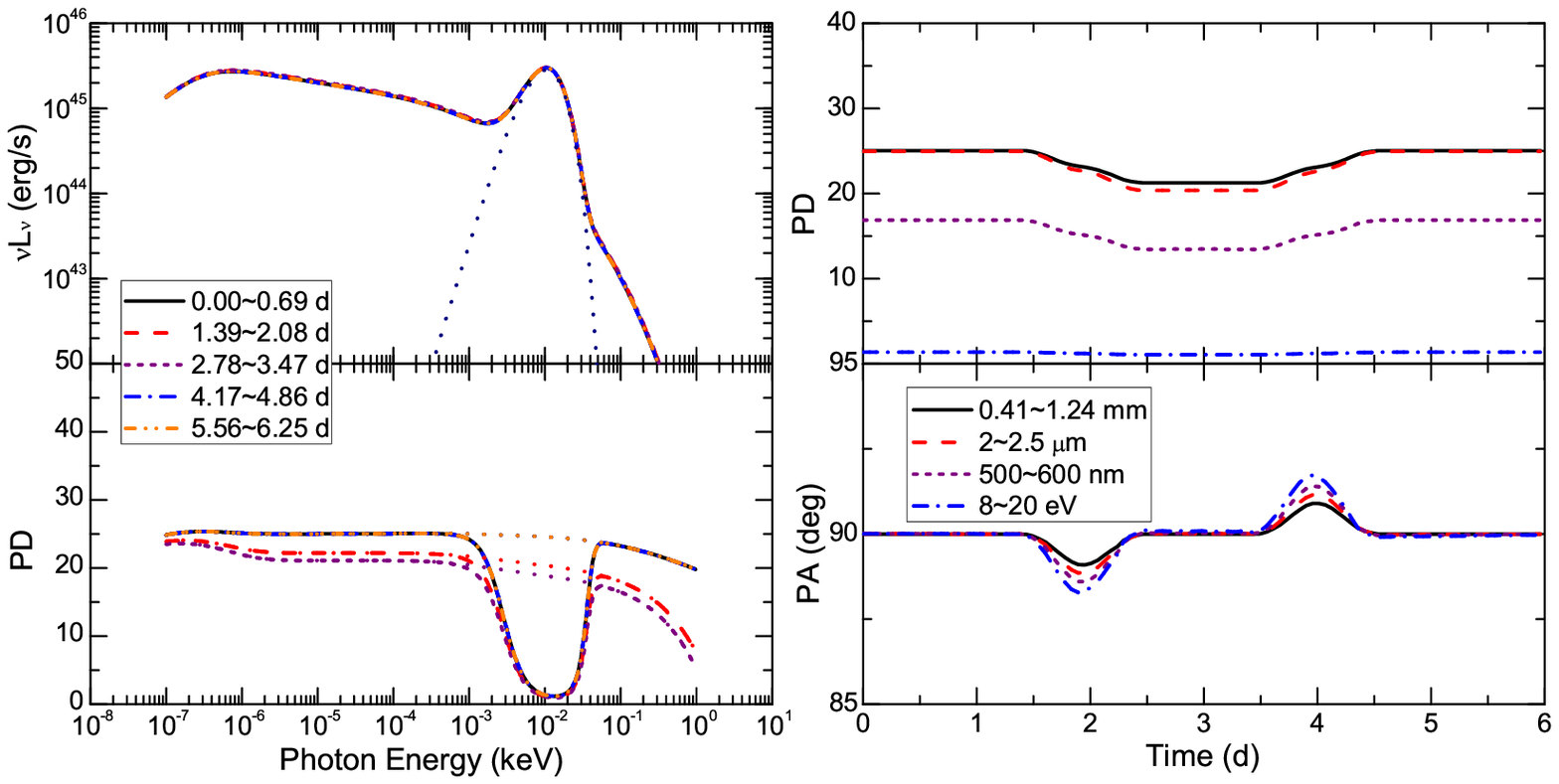}
\caption{Case 4: Moderate change of the magnetic field with Geometry III.
Panels and line styles are as in Fig. \ref{Case11}.
\label{Case13}}
\end{figure}

We present one case for the combined geometric effect in Fig. \ref{Case13}. This time we apply 
a moderate change in the magnetic orientation and strength during the presence of the shock,
with the same volume of the emission region as in Geometry I ($R:Z=1:\frac{4}{3}$),
but now the shock will not occupy the whole layer of the emission region, but only the center of it,
so that the region that will be affected by the shock, has $A:Z=\frac{2}{3}:3$, (hereafter Geometry III).
The current situation is equivalent to a Geometry II case but surrounded by a large quiescent region. 
Hence the positive $q$ in the
shock region will not be able to dominate over the negative $q$ in the quiescent region. Consequently, 
we find a similar behavior as in Case 3,
where the $PA$ rotates back to $90^{\circ}$ and exhibits a step phase. The difference is that the $PD$ 
slightly decreases during the flare peak,
since the shock will give a boost to the toroidal contribution to the polarization, diminishing part of 
the initial poloidal contribution.

\section{Application to 3C279}

Based on the insights gathered in the above study and Paper I, we now present the first simultaneous 
fitting of snapshot SEDs, multifrequency light curves, and time-dependent polarization signatures of 
a blazar, using the example of a flare of 3C279.
The FSRQ 3C279, located at a moderate redshift of $z = 0.536$, is one of the most well-observed members 
of the blazar class. It caught the attention of the high-energy astrophysics community due to its very 
bright gamma-ray flaring at the beginning of the Compton Gamma-Ray Observatory mission in the early 
1990s, and has since then been the target of many dedicated multi-wavelength observing campaigns 
\citep{Abdo10,Hartman96,Maraschi94,Wehrle98}. It is bright and variable on a large range of 
time-scales, across the electromagnetic spectrum, from radio through gamma-rays, and is one of 
only three FSRQs detected in very-high-energy gamma-rays (i.e., photon energies of $E > 100 GeV$) 
by the MAGIC telescope \citep{Albert08}.

\begin{figure}[ht]
\centering
\includegraphics[width=0.5\textwidth]{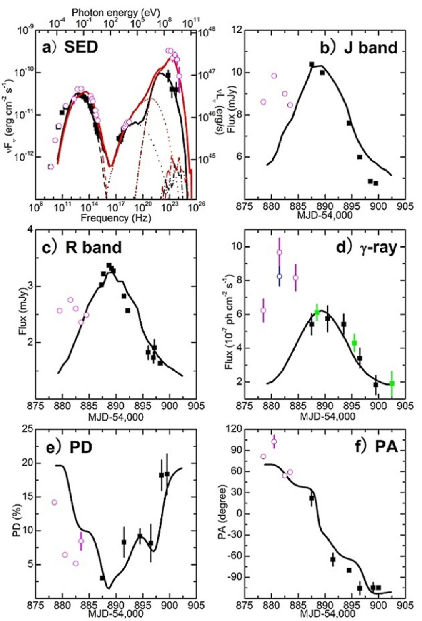}
\caption{Data and model fits to multi-wavelength SEDs, light curves and polarization signatures 
during the second flare of 3C279 in 2009. Data are from \cite{Abdo10,Hayashida12}. Hollow (filled) 
data points refer to the first (second) flare. a) SED, black squares are from Period E in \cite{Hayashida12} 
(MJD 54897 -- 54900), corresponding to the end of the second flare. The black curve is the model SED from 
the simulation at the same period; the red curve is the simulated SED at the peak of the flare. Hollow 
magenta circles are from Period D (MJD 54880 -- 54885), corresponding to the end of the first flare. b,c) 
J and R Band flux: black squares are from the period of the second flare, hollow magenta circles are from 
the first flare; the black curves are the simulated light curves. d) Gamma-ray photon flux light curve: 
black squares and hollow magenta circles are 3-day averaged data from the second and the first flare, 
respectively, while green squares and hollow blue circles are the 7-day averaged data, respectively. e, f) 
$PD$ and $PA$ vs. time: black squares and hollow magenta circles are from the second and the first flare, 
respectively; curves are the simulated polarization signatures.
\label{Fitting}}
\end{figure}

During an active (i.e., high-flux) phase from Nov. 2008 to Mar. 2009, 3C279 exhibited multi-wavelength 
flaring activity, with a period of optical polarization variations lasting $\sim 20$ days \citep{Abdo10}. 
The multi-wavelength light curve data are displayed in Fig. \ref{Fitting}. They suggest that this flaring 
episode is actually composed of two sequential flares. The first flare dominates around MJD 54880, where 
a sudden doubling in gamma-ray flux is accompanied by a relatively small infrared-to-optical flare and 
rather erratic changes of the $PD$ and $PA$. The second flare, which dominates the last $\sim 15$ days, 
constitutes a correlated flare of the infrared-to-optical and gamma-ray emissions, during which the $PD$ 
drops to nearly zero and then recovers to $\sim20\%$, accompanied by a $\sim180^{\circ}$ rotation of 
the $PA$. This was the first time that such a clear correlation between optical/gamma-ray flaring and 
a $PA$ rotation was observed. Based on the apparently time-symmetric profile of the second flare, we 
suggest that it actually lasted $\sim20$ days, but is overwhelmed by the decaying phase of the first 
flare for the first $\sim5$ days. Here we present a consistent interpretation of the spectral energy 
distribution (SED), multi-wavelength light curves, and time-dependent polarization features, for the 
second flare.

Table \ref{table} lists the most relevant parameters in our simulation that yields the best fit to the 
data of 3C279, and the results are shown, in comparison with the data, in Fig. \ref{Fitting}. The 
low-frequency component of the SED is dominated by synchrotron emission from nonthermal electrons, 
along with a weak thermal component from the central accretion disk around the black hole powering 
the relativistic jet. We assume a leptonic origin for the high-energy component of the SED, which 
is composed of an SSC contribution, dominating from X-rays to soft gamma-rays, and an EC contribution 
which dominates the emission in the Fermi ($\sim GeV$) range. We also assume that there exists a 
turbulent magnetic field component, which is initially uniform everywhere. We assume the disturbance 
to be associated with a magnetic energy dissipation process, e.g., magnetic reconnection, instead of a 
shock. It has been suggested that magnetic reconnection can dissipate a large fraction of the magnetic 
energy to produce a nonthermal power-law distribution of relativistic particles \citep{Guo14}. A 
possible underlying physical picture is that on the trajectory of the emission region, it encounters 
a flat stationary region, where the poloidal component of the helical magnetic field is in the opposite 
direction of that in the emission region. Therefore the poloidal component will be dissipated during 
the presence of the disturbance, due to magnetic reconnection, leading to particle acceleration (we 
use a simple injection to mimic this effect) and a locally stronger turbulent magnetic-field component 
\citep{Daughton11,Guo14}.

The time-dependence of the polarization signatures is similar to Case 1 of the previous section. 
Before interacting with the disturbance, the entire emission region contributes uniformly, and therefore, 
due to the axisymmetry of the underlying geometry and the mildly stronger poloidal component 
($\theta_B = 33^{\circ}$) in the quiescent state, the polarization is dominated by the poloidal 
contribution. When the interaction with the disturbance starts (red in Fig. \ref{Sketch}), we assume 
that the poloidal component in this area decreases, but the toroidal component remains nearly unaffected, 
hence the quiescent-state polarization will be gradually canceled out by the flaring toroidal contribution. 
As a result, the $PD$ drops and the $PA$ starts to rotate towards the dominant toroidal direction. As the 
disturbance moves forward, the flaring region becomes larger, resulting in increasing fluxes in both the 
synchrotron and the Compton emissions. Towards the end of the interaction period, the flaring region 
(green in Fig. \ref{Sketch}) reaches its maximum size, leading to the observed flux maximum. Meanwhile, 
the magnetic field throughout the green flaring region in Fig. \ref{Sketch} has approximately equal 
contributions from the near and far sides of the cylinder, mimicking the initial axisymmetry, except 
for a mildly stronger toroidal magnetic-field contribution ($\theta_B=62^{\circ}$). This flaring toroidal 
contribution is just strong enough to dominate over the poloidal contribution in the quiescent state. 
Consequently, the $PD$ drops to nearly zero, while the $PA$ reflects the dominant toroidal magnetic-field 
direction. After this flux peak, the flaring region gradually becomes smaller and moves to the upper-left 
(blue in Fig. \ref{Sketch}), and the light curves and $PD$ recover to their initial states with approximately 
time-symmetric profiles. However, since the toroidal component on the far side of the cylinder is opposite 
to that on the near side, the $PA$ instead completes a $\sim 180^{\circ}$ swing to a direction that is 
indistinguishable from its initial position due to the $180^{\circ}$ ambiguity of the $PA$.

\begin{table}[ht]
\scriptsize
\parbox{1.0\linewidth}{
\centering
\begin{tabular}{|l|c|c|}\hline
Parameters                                          & \multicolumn{2}{c |}{General Properties}   \\ \hline
Bulk Lorentz factor $\Gamma$                        & \multicolumn{2}{c |}{20}                   \\ \hline
Length of the emission region $Z$ $(10^{17}cm)$     & \multicolumn{2}{c |}{4.8}                  \\ \hline
Radius of the emission region $R$ $(10^{17}cm)$     & \multicolumn{2}{c |}{2.4}                  \\ \hline
Length of the disturbance $L$ $(10^{17}cm)$         & \multicolumn{2}{c |}{1.28}                 \\ \hline
Orientation of LOS $\theta_{\rm obs}$ $(^{\circ})$      & \multicolumn{2}{c |}{90}                   \\ \hline
Electron acceleration time-scale $t_{\rm acc}$ $(Z/c)$  & \multicolumn{2}{c |}{0.36}                 \\ \hline
Electron escape time-scale $t_{\rm esc}$ $(Z/c)$        & \multicolumn{2}{c |}{0.062}                \\ \hline
Electron background temperature $T$ $(m_ec^2/k_B)$  & \multicolumn{2}{c |}{100}                  \\ \hline
Parameters                                          & Quiescent         & Active                 \\ \hline
Total magnetic field strength $B$ $(10^{-2}G)$      & $7.2$             & $5.4$                  \\ \hline
Helical magnetic strength, $B_H$ $(10^{-2}G)$       & $4.8$             & $2.6$                  \\ \hline
Helical pitch angle $\theta_{B}$ $(^{\circ})$       & $33$              & $62$                   \\ \hline
Initial electron density $n_e$ $(cm^{-3})$          & $5.5$             & --                     \\ \hline
Injected electron minimum gamma $\gamma_{\rm min,inj}$  & --                & $2000$                 \\ \hline
Injected electron maximum gamma $\gamma_{\rm max,inj}$  & --                & $4000$                 \\ \hline
Injected electron power-law index $p_{\rm inj}$         & --                & $4$                    \\ \hline
Injection rate $Q_{\rm inj}$ $(10^{43} erg*s^{-1})$          & --                & $5$                    \\ \hline
\multicolumn{3}{c}{}\\
\end{tabular}}
\caption{Summary of parameters. The parameters are defined in the same way as in Section 3, except for a few 
differences. In this fitting, we assume the radius of the disturbance is equal to that of the emission region, 
so that $A=R$. As we mentioned in Section 3, the initial electron distribution will evolve to an equilibrium 
according to the Fokker-Planck equation, before interacting with the disturbance, therefore we only list the 
initial electron density $n_e$, as the spectrum will be determined by the Fokker-Planck equation. We assume 
that the escaped electrons will be balanced by the electrons picked up from the thermal background, with a 
temperature $T$. The magnetic field has two components, a helical component $B_H$ along with a turbulence; 
the total magnetic field strength will be $B$. The disturbance will change the layers in the emission region 
at its location into an active state. In such situation, the magnetic field energy will be dissipated, so 
that both the total strength and the helical strength will decrease and the pitch angle will alter. The 
dissipated energy will become the energy resource for particle acceleration, which we handle it by an 
injection at the front of the disturbance with an energy injection rate $Q_{inj}$ into the emission region, 
with minimum and maximum Lorentz factors $\gamma_{min,inj}$ and $\gamma_{max,inj}$, and power-law index 
$p_{inj}$. Both the initial and the injected electrons will evolve according to the Fokker-Planck 
equation.\label{table}}
\end{table}

\section{Discussions}

Polarization signatures are known to be highly variable, and $\ge 180^{\circ}$ polarization angle swings are 
frequently observed \citep[e.g.][]{Larionov13,Morozova14}. Generally, the observed $\ge 180^{\circ}$ $PA$ 
swings are accompanied by one or several sequential apparently symmetric $PD$ patterns which drops from an 
initial value to zero then reverts back. In addition, both the $PD$ and $PA$ patterns appear to be smooth. 
Several mechanisms have been proposed to interpret the $PA$ rotations, such as an emission region moving 
along a curved trajectory or a bending jet, or streamlines following helical magnetic-field lines, or 
stochastic activation of individual zones in a turbulent jet \citep{Abdo10,Marscher10,Marscher14}. While 
they may have their own virtue in understanding polarization variations, none of these models has so far 
been able to explain spectral variability properties, symmetric light-curve profiles, and correlated 
symmetric polarization variability features in one coherent model. On the other hand, the LTTEs coupled 
with an axisymmetric emission region in our model naturally explain the apparently time-symmetric features 
of multi-wavelength light curves and polarization variations and their intrinsic correlations, which also 
appears to be the simplest model with the smallest number of fine-tunable parameters.

Based on our model, the continuous $180^{\circ}$ $PA$ swings will pose some serious constraints on the 
physical background of the emission region, namely: the pitch angles of the quiescent and the active states 
shall be on each side of a critical point $\theta_c$, a moderate strength of flare, a moderate change of 
the pitch angle of the helical magnetic field, and a geometry of the emission region where $Z\gtrsim R$. 
If the pitch angles of the quiescent state and the active state are both within the range of $0$ to 
$\theta_c$ or $\theta_c$ to $90^{\circ}$, where in our case $\theta_c\sim 55^{\circ}$, then the projected 
magnetic field will always provide the total Stokes parameters in the form of $1,-\vert q \vert, u$ or 
$1,\vert q \vert, u$, respectively, resulting in the $PA$ fluctuating around $270^{\circ}$ or $180^{\circ}$, 
respectively. Additionally, if the flare amplitude or the pitch angle change is too weak, then throughout 
the flaring activity the total polarization signatures will be dominated by the quiescent state, therefore 
$PA$ will not complete a $180^{\circ}$ swing. While if on the other hand they are too strong, then the 
active region will dominate the polarization signatures for a much longer time, giving rise to a step-like 
feature as shown in Paper I. Notice that, however, this feature is different from the ``step-phase'' feature 
introduced by a $Z>2R$ geometry (Case 2), because the step-like feature caused by strong variations can only 
reach $180^{\circ}$ at the very peak of the flare, where the $u$ component of the Stokes parameters can be 
fully canceled out. On the other hand, the ``step-phase'' due to the geometry can be diagnosed in the 
observation as completely flat \citep[][]{Ikejiri11}.

In addition to the above constraints, a simultaneous fitting of all spectral, light curve, and polarization 
properties can exclude a wide range of possible scenarios.
Therefore, our fitting results place unprecedented constraints, most of which do not depend on the details 
of the model. Since in a leptonic model interpretation, the gamma-rays from 3C279 are produced by the EC 
process, an excess of either external photon field or nonthermal electrons is necessary to produce a gamma-ray 
flare. However, the former fails to produce the data, since such excess is unlikely to generate a correlated, 
time-symmetric multi-wavelength flare, but instead may lead to an anti-correlated behavior in the synchrotron 
emission due to excess radiative cooling. Also, the infrared-to-optical flare amplitude is smaller than that 
of the gamma-ray flare, and the X-ray emission, which represents the low-energy end of the SSC emission, shows 
almost no variability. This implies that the total magnetic field strength has to decrease during the flare. 
We suggest that the dissipated magnetic energy can be converted into nonthermal particle energy through 
magnetic reconnection. Most importantly, as we have discussed above, a smooth $PA$ rotation and the nearly 
zero $PD$ at the flare peak (green in Fig. \ref{Sketch}) imply that the toroidal magnetic-field component 
should be mildly stronger than the poloidal one in the active state. However, since the excess of nonthermal 
electrons and the amount of the magnetic field alteration, which are linked by the reconnection process, 
have been well constrained, the poloidal and toroidal B-field components should remain comparable throughout 
the process. Moreover, the relative percentage of the turbulent magnetic-field contribution increases in the 
active state. Nevertheless, we notice that, just like most multiwavelength models of blazar emission, our 
model underestimates the radio flux, and also overestimates the $PD$ there. This is because the radio 
emission is known to be produced on larger scales, on which the magnetic field appears more disordered. 
In addition, the environment in which our emission region is located is still opaque to radio emission.

Models of relativistic shocks propagating through the jet have been widely used to explain blazar flaring 
activities \citep{Boettcher10,Joshi07,Marscher85,Spada01}. Such models naturally provide excess nonthermal 
electrons. However, they are expected to compress the magnetic field, leading to a change in $\theta_B$ with 
an increase in the overall magnetic-field strength. Consequently, shocks are unlikely to fit our constraints. 
Alternatively, the fitting constraints strongly favor a magnetic energy dissipation process during the flare. 
We find in our simulation that the dissipated magnetic energy during the disturbance is comparable to the 
necessary amount of particle energy increase required to generates the flare. Simulations of magnetic energy 
dissipation have demonstrated that the energy stored in magnetic shear can be efficiently converted into a 
power-law distribution of relativistic particles \citep{Guo14}. This process will reduce the magnetic field 
component that is subject to dissipation and can therefore change the magnetic-field pitch angle. Moreover, 
the magnetic-field topology inside the dissipation zone is likely to become turbulent, thus it will strengthen 
the turbulent magnetic-field contribution \citep{Daughton11}.

For the smaller fluctuations in the polarization signatures (generally $PA$ varies by $<90^{\circ}$), 
although some apparently symmetric profiles can be noticed, the general patterns appear very complicated 
\citep[][]{Jorstad13, Covino15}. In addition, those fluctuations often happen during the quiescent states, 
with lower $PD$. This implies that some inhomogeneity and more complex geometry, or some turbulence are 
required \citep{Marscher14}.

We suggest that, if a step phase along with time-symmetric
variability of the polarization features is observed, especially in the $PA$ profile, as it will be 
less affected by a possible turbulent B-field component, this may serve as a powerful constraint on 
the size and the geometry of the emission region. Since the step phase will serve as a measure of the 
length of $Z-2R$, while the two symmetric $PA$ ($PD$) alterations will imply the length of $A$. In this 
way, the size and the geometry of the region that is affected by the disturbance can be constrained, 
while the size of the entire emission region can then be estimated by the flux and polarization percentage.

\acknowledgments{We thank the anonymous referee for a careful review and very helpful suggestions to 
improve the paper. HZ, FG and HL are supported by the LANL/LDRD program and by DoE/Office of Fusion 
Energy Science through CMSO. XC acknowledges support by the Helmholtz Alliance for Astroparticle 
Physics HAP funded by the Initiative and Networking Fund of the Helmholtz Association. MB acknowledges 
support by the South African Research Chairs Initiative (SARChI) of the Department of Science and 
Technology and the National Research Foundation\footnote{Any opinion, finding and conclusion or 
recommendation expressed in this material is that of the authors and the NRF does not accept any 
liability in this regard.} of South Africa. Simulations were conducted on LANL's Institutional 
Computing machines.}

\clearpage

\end{document}